# Using Single-Trial Representational Similarity Analysis with EEG to track semantic similarity in emotional word processing

Feng Cheng, Tufts University

**Abstract.** Electroencephalography (EEG) is a powerful non-invasive brain imaging technique with a high temporal resolution that has seen extensive use across multiple areas of cognitive science research. This thesis adapts representational similarity analysis (RSA) to single-trial EEG datasets and introduces its principles to EEG researchers unfamiliar with multivariate analyses. We have two separate aims: 1) we want to explore the effectiveness of single-trial RSA on EEG datasets; 2) we want to utilize single-trial RSA and computational semantic models to investigate the role of semantic meaning in emotional word processing. We report two primary findings: 1) single-trial RSA on EEG datasets can produce meaningful and interpretable results given a high number of trials and subjects; 2) single-trial RSA reveals that emotional processing in the 500-800ms time window is associated with additional semantic analysis.

# Introduction

Electroencephalography (EEG) is a non-invasive brain imaging technique with high temporal resolution. EEG records electrical activities on the scalp through recording channels (electrodes), and EEG signals are thought to reflect excitatory and inhibitory postsynaptic potentials in populations of cortical neurons. EEG is typically employed by studies that investigate transient neurocognitive processes that are inaccessible to other imaging techniques such as functional Magnetic Resonance Imaging (fMRI).

Representational similarity analysis (RSA, Kriegeskorte et al., 2008) is a powerful multivariate analysis that investigates patterns of neural activation. Since its introduction, RSA was widely adapted by fMRI studies and yielded novel results complimenting findings from univariate analyses. Despite its capability, however, RSA is seldom employed in EEG studies. In this thesis, we want to introduce RSA to EEG researchers unfamiliar with multivariate analysis. In particular, we will explain the basic principles behind RSA and present its methodological adaptation to EEG datasets. To demonstrate the power of RSA, we will apply this analysis to a single-trial EEG dataset and investigate the role of semantic meaning in emotional word processing.

## Univariate and Multivariate analyses

In general, univariate analyses are primarily concerned with the degrees of activities of a group of recording channels. In a typical univariate EEG study, the recordings of trials are segmented into equal-sized time-locked and phase-locked evoked responses called epochs. Epochs belonging to the same experimental condition are then averaged to obtain an event-related potential (ERP). Regular waveforms observed in an ERP segment are referred to as components and are characterized by their polarities, latencies, and locations on the scalp. These components are taken to be the signatures of the underlying neurocognitive processes triggered by a type of stimuli.

While the univariate analyses treat channels as independent measures and selectively explore their overall *degrees of activations*, the multivariate analyses investigate the ensemble of activation levels of a large group of channels by analyzing their *patterns of activations* (Grootswagers et al., 2017). Multivariate analyses can hence utilize finer-grained information that is inaccessible to traditional univariate analyses. Two approaches dominate the multivariate analyses in brain imaging studies of cognitive science. While the *Decoding* approach predicts condition/stimulus information from the activation patterns of recording channels, the *Encoding* approach predicts activation patterns from condition/stimulus information (Kriegeskorte & Douglas, 2019).

The difference between univariate and multivariate analysis is usually understood as such: univariate analyses reveal the onset of a particular neurocognitive process, whereas multivariate analyses reveal the representational content encoded in these processes. However, this direct contrast between process and representation is misleading. Strictly speaking, patterns of neural activities at any level can all be interpreted as the neural representation of a stimulus (Dennett, 1987). Processes and representations are tightly linked with each other since the onset of a neurocognitive process determines the kind of neural representation encoded in the brain. Consider, for example, that a neurocognitive process is specialized in analyzing emotional stimuli. The emotional content of the stimuli will only become salient in the brain when and after this neurocognitive process is triggered. Therefore, the distinction between univariate and multivariate is more clearly defined in their difference in sensitivities (coarse vs. fine-grained) of the recorded neural signal and the nature of the neural response (activation vs. pattern) each approaches investigate (Grootswagers et al., 2017).

### Adapting RSA to EEG datasets

Representational Similarity Analysis (Kriegeskorte et al., 2008) falls into the category of encoding analysis because it predicts neural data from condition/stimulus information. But instead of predicting the activation levels of an ensemble of recoding channels, RSA predicts the summary statistics of the response patterns elicited by all experimental stimuli in the form of Representational Dissimilarity Matrix (RDM; Kriegeskorte & Diedrichsen, 2019). Specifically, A neural RDM is a square matrix defined by the pairwise discriminability (or dissimilarity) values of all pairs of stimulus-specific channel activation patterns. The discriminability (or dissimilarity) between two activation patterns is typically measured through their correlation distance (1 - correlation coefficient). On a conceptual level, the RDM projects the neural responses of all stimuli into a high dimensional multivariate response space, and the discriminability of any two stimuli reflects the geometrical distance of their projections in this space (Nili et al., 2014).

Critically, an RDM can be calculated from any variable that measures the stimulus-specific properties (hereby referred to as a variable RDM) if a valid dissimilarity function can be defined. For example, consider the valence ratings of a set of word stimuli: a valence RDM can be obtained by calculating the absolute differences between all pairs of valence ratings of the words in the set. Since the representational geometries captured by RDMs only retain the second-order relationships between all stimulus-specific data, RDMs can be compared to probe overlapping second-order relationships. This comparison process is typically performed by measuring the correlation coefficient of the upper triangular parts of two RDMs (Kriegeskorte et al., 2008). Intuitively, a high coefficient value between the two RDMs entails a high level of correspondence in pairwise dissimilarity values, and therefore a high degree of overlap between the two representational geometries projected onto a common representational space (Popal et al., 2019). Hence the resultant correlation coefficient from RSA reveals the degree of overlapping representational content encoded by various forms of stimulus-specific data.

RSA was traditionally applied to fMRI data, although some recent M/EEG studies incorporated RSA to study the recorded spatial activation patterns across the recording time points (e.g., Cichy et al., 2014, Kaneshiro et al., 2015). Nonetheless, RSA studies are still relatively scarce in the M/EEG literature. One major worry that prevents M/EEG researchers from applying this analysis concerns the nature of the spatial activation patterns recorded in the M/EEG dataset. Specifically, the spatial activation distribution recorded by M/EEG channels does not resemble the spatial activation distribution on the cortex. This is known as the source ambiguity problem. This problem is especially pronounced in EEG since the cortical voltage distribution is significantly distorted as it travels through the scalp (Luck, 2014). Although it is possible to estimate the source locations from M/EEG recordings, their actual locations still remain ambiguous since there is no unique solution (Stokes et al., 2015). It is therefore suspected that the smeared spatial patterns picked up by the recording channels are no longer meaningful. In addition to this worry, the low density of recording channels of M/EEG also raises concerns regarding the validity of spatial activation patterns recorded.

This worry about source ambiguity is valid if RSA is used to reveal the *spatial distribution* of representational content encoded in various brain regions. And indeed, RSA studies using fMRI data are primarily interested in this spatial distribution: they usually confine RSA to incorporate the activation values of voxels in an anatomically defined cortical area to study the representational content encoded in this substrate. However, RSA studies using M/EEG datasets are primarily interested in the *temporal evolution* of the representational content. These studies utilize the activation values of all channels at a time point to investigate the representational content encoded in the activation pattern of the entire scalp. The analysis is then repeated for all time points to reveal the evolution in the representational contents over time.

Because all the recording channels are incorporated to yield a spatial activation pattern at each single time point, the source ambiguity problem will not affect RSA. As long as different stimuli trigger different spatial activation patterns picked up by the recording channels, the neural RDM can successfully create a meaningful representational geometry even when the true source of these patterns cannot be determined (Cichy & Pantazis, 2017). In support of this insight, studies show consistent successes in decoding stimulus information from M/EEG data (Wolff et al., 2015; Grootswagers et al., 2017). Further, the low density of the recording channel is not a significant problem since a robust performance in decoding accuracy can still be maintained when only 32 EEG channels are employed (Cichy & Pantazis, 2017, Wang et al. 2020).

These previous studies address the worry mentioned above and provide support for applying RSA to M/EEG datasets. In this thesis, we want to explore the use of RSA on single-trial EEG datasets, and we aim to introduce this analysis to M/EEG researchers unfamiliar with multivariate analysis.

## Capturing Semantic Meaning using Distributed Semantic Models

Semantic meaning is notoriously hard to capture. For example, competent speakers of a language often experience great difficulties in defining the words they have no trouble using (Miller & Charles, 1991). In this thesis, we will explore an approach that operationalizes semantic meaning by focusing on the *semantic similarity/dissimilarity* of a group of words. The concept of semantic similarity is very natural to the human mind, and studies found that subjects' similarity judgments are consistent and retain high inter-subject agreements (Miller & Charles, 1991, De Deyne et al., 2016). While many studies directly use the subject's explicit judgements to define semantic similarity, some propose computational techniques or models that estimate semantic similarity from other measures. For example, similarity functions such as Wu Palmer similarity (Wu & Palmer, 1994) can be applied to the hierarchical linguistic organizations of words recorded in the Wordnet database (Miller, 1998) to calculate semantic similarity. De Deyne et al. (2016) recently proposed a graph model that utilizes a large database of subjects' responses in a semantic association task to capture general semantic dissimilarity relationships between words.

A particularly successful type of model that is often employed by the latter group of studies is the Distributional Semantic Model (DSM). This type of model is constructed by machine learning algorithms that calculate semantic similarities of words entirely from their distributional statistics across an extensive database of texts. The root of DSM can be traced back to the *distributional hypothesis* in linguistics that is often attributed to Harris (1954). In short, this hypothesis states that words with similar meanings tend to occur in similar contexts, and the semantic meaning of every single word can be captured by a long word vector representing the encounters with a word across all linguistic contexts in the entire corpus (Miller & Charles, 1991). Under this view, the semantic similarity of two words can be directly calculated by comparing their word vectors, and similar word vector compositions entail similar distribution patterns across contexts.

Different DSM models differ greatly in their definitions of the linguistic context and their formulations of the word vectors. Older DSMs such as HAL (Lund & Burgess, 1996) and LSA (Landauer & Dumais, 1997) harness global co-occurrence statistics over large context windows to produce word vectors. Newer DSMs such as word2vec (Mikolov et al., 2013) utilize feed-forward neural networks to predict words given a small context window (or vice versa). A word vector is defined as the word's distributed activation pattern in the network's internal projection layer that efficiently summarizes the word's distributional statistics across contexts. The semantic meaning captured by the newer models such as word2vec is highly sophisticated. For example, the word2vec word vectors demonstrate some compositional structures like "king - man + woman ≈ Queen" (Mikolov et al., 2013 June).

It is important to note that these DSMs are only trained on linguistic data and essentially capture the "meaning in the text" (Sahlgren, 2008). Hence, a common criticism of these models is that they do not capture the rich extra-linguistic content contained in semantic meaning (Hagoort et al., 2004) and cannot serve as a model of semantic meaning representation in the brain. However, this criticism ignores the fact that our languages are not independent of the world, and rich extra-linguistic information is often implicitly encoded in linguistic information. Studies have shown that DSMs capture extra-linguistic information such as the vertical location of objects in the world, perceptual modalities, and affective dimensions (for review, see Günther et al., 2019). They can also accurately predict human word similarity judgments (Baroni et al., 2014). Although the exact kind of information captured by DSM is a topic of current research, it is safe to conclude that it is one solid way to model semantic representation in the human brain.

## Connecting DSM semantic dissimilarity with neurocognitive processes

A semantic RDM of a group of words can be easily constructed by using their associated DSM-based word vectors. If two words have similar word vectors, then they occur in similar contexts and are semantically similar according to the distributional hypothesis. A semantic RDM of a set of words can be constructed by comparing the cosine dissimilarity (Pennington et al., 2014) between all pairs of word vectors. In principle, this semantic RDM can then be compared with neural RDMs to investigate whether and what semantic information is encoded by the neural activation patterns.

Many multivariate studies have successful connected DSM word vectors with neural activity. For example, Lyu et al. (2019) successfully applied RSA to the EMEG recording and showed that the semantic RDM obtained from DSM word vectors could successfully predict the neural RDM. In the same vein, both the encoding (Mitchell et al., 2008) and decoding analyses (Wehbe et al., 2014, Devereux et al., 2010, Pereira et al., 2018) were highly successful in mapping the neural response patterns of a word to its associated DSM word vector (or vice versa) in the context of language processing. Together, these studies provide ample evidence supporting the use of DSM-based word vectors in modeling neural representation of semantic meaning.

RSA's ability to model neural semantic processing with DSM opens up novel opportunities to investigate psycholinguistics processes related to semantic meaning. In this thesis, we will explore the role of semantic meaning in emotional word processing. ERP literature indicates that neurocognitive processes in the 500-800ms time window after stimulus onset is related to emotional processing. In particular, the ERP component *Late Positivity Potential* (LPP) is reported to index emotional processing in this time window. The amplitude of LPP reflects the stimulus's motivational significance — the extent to which a stimulus activates an appetitive or aversive motivational system in the brain (for review, see Hajcak & Foti, 2020). Emotional stimuli are naturally significant, and an increase in LPP amplitude reflects a relatively automatic and sustained engagement with the emotionally significant content since it might encode critical information about potential threats or opportunities. Additionally, LPP is found to be highly sensitive to context and can be modulated by the experimental task (Delaney-Busch et al., 2016; Fields & Kuperberg, 2016). For example, the Delaney-Busch et al. 2016 EEG study reported that the words' valence category has no effect on LPP when the words are presented without a task. However, when the subjects are explicitly asked to classify the stimulus word into one of the three valence categories (pleasant/positive, neutral, unpleasant/negative), both the positive and the negative words elicit a larger LPP response relative to neutral words.

Although the univariate literature on the LPP component reveals that motivationally significant stimulus can trigger emotional processing in the 500-800ms time window, little is known about the exact content being processed. One possibility we want to explore here is that the neurocognitive processes in this time window are associated with additional semantic analysis of the motivationally significant stimuli.

## The Present Thesis

The present thesis has two separate but related aims. Our first aim is to explore the effectiveness of RSA on single-trial EEG datasets. In the context of RSA, single-trial datasets refer to those with unique stimuli for each trial. Single-trial EEG datasets are very common in psycholinguistics studies. Because human language cognition is very sensitive to the local context, repeating stimulus almost always significantly alters the stimulus-specific neural responses due to semantic priming or repetition suppression (for their effects on the N400 component, see the review of Kutas & Federmeier, 2011). Single-trial EEG datasets pose a unique challenge for RSA. Individual recordings are noisy in all recording techniques, and many RSA studies usually employ the conditional-rich design to raise the signal-to-noise ratio (Kriegeskorte et al., 2008). In this design, each stimulus is repeated in several trials, and the neural responses elicited in these trials are averaged to yield one stimulus-specific neural recoding. The neural RDM is then constructed from these average recordings. However, the use of unique stimuli means that the neural RDM can only be constructed from noisy single-trial recordings. Since single-trial EEG recordings are perhaps the noisiest data among all types of neural recordings, there is a genuine worry about the utility of RSA on single-trial EEG datasets. It is suspected that RDM built from single-trial EEG data might not capture a meaningful neural representational geometry. In this thesis, we will test the effectiveness of RSA on the *Kiloword* dataset reported by Dufau et al. (2015). In the study, 960 unique nouns are presented to the subjects while they performed a semantic categorization task. We will investigate the effect of the number of trials and subjects on the RSA result. We will also explore three interpolation approaches to experiment techniques that can potentially increase the signal-to-noise ratio of the RSA results.

Our second aim is to investigate the role of semantic meaning in emotional processing. We hypothesize that the neurocognitive processes in the 500-800ms time window are associated with additional semantic analysis of the motivationally significant stimuli. To test this hypothesis, we will reanalyze the *EmSingle* dataset reported by Delaney-Busch et al. (2016) with RSA. In the study, 468 unique nouns are presented to the subjects while they performed the valence judgment task. This study reported that valenced words (negative & positive) triggered a robust LPP effect in the 500-800ms time window. In this thesis, we will model the neural representation of semantic meaning using the semantic RDM built from word2vec, and we will compare the semantic RDM with the neural RDMs in the 500-800ms time window to probe overlapping representational geometry. Our hypothesis leads to two predictions about the RSA results: 1) The semantic RDM should be significantly correlated with the neural RDM in the 500-800ms time window; 2) After confining RSA to the trials in each of the three valence conditions (negative, neutral, positive), the semantic RDM should be significantly correlated with the neural RDM in the 500-800ms time window in the positive and the negative condition, but not in the neutral condition. Further, we are also curious about the content of the semantic meaning captured by the semantic RDM. To this end, we analyze the semantic RDM built from the EmSingle stimulus set with multidimensional scaling (MDS, Kruskal, 1978) and Hierarchical Clustering (Johnson, 1967) and investigate whether and to what extent common psycholinguistics variables are encoded in this RDM.

# Methods

In this section, we will first break down the principles of RSA to introduce this analysis to EEG researchers. We will then explain how we test the effectiveness of RSA on single-trial EEG data using the Kiloword dataset (Aim 1). Finally, we will present the methodological details of our analyses of the EmSingle dataset (Aim 2).

## Basic Principles of RSA

### Calculating RDM

The first step in RSA is to calculate the variable and neural RDMs. Since we are applying RSA to single-trial data, we will use "trial" and "stimulus" interchangeably. Conceptually, an RDM captures the representational geometry of a variable of interest, and multiple RDMs that share a common stimulus set can be compared with each other through RSA to quantify overlapping representational content. In the context of psycholinguistics studies, we are primarily interested in using RDMs of relevant psycholinguistic variables to predict the neural RDMs. A variable RDM can be obtained by calculating the pairwise dissimilarity values of all possible pairs of stimuli (Fig 1a). For the common psycholinguistics variables (e.g., word length, concreteness (Brysbaert et al., 2014), emotional valence and arousal (Warriner et al. 2013)), the pairwise dissimilarity values of words can be simply defined as the absolute difference in their values. For vector-based semantic variables (e.g., word2vec embeddings), the pairwise dissimilarity value can be defined as cosine dissimilarity between the word vectors. Notably, some psycholinguistics properties such as Orthographic Levenshtein Distance (Yarkoni et al., 2008) are naturally defined in pairwise relationships and can be used to compute the associated RDMs directly.

The process for calculating neural RDM demands extra attention. In typical EEG studies, the continuous EEG recordings are segmented into equal-sized epochs (trials) that are time-locked to the stimuli onsets. The recording data of each trial can be stored as a matrix, with rows representing channels and columns representing time points. In the naive approach, we can only focus on the spatial activation patterns of all channels at time point $t$. We will represent each trial's spatial activation pattern at time $t$ with a vector

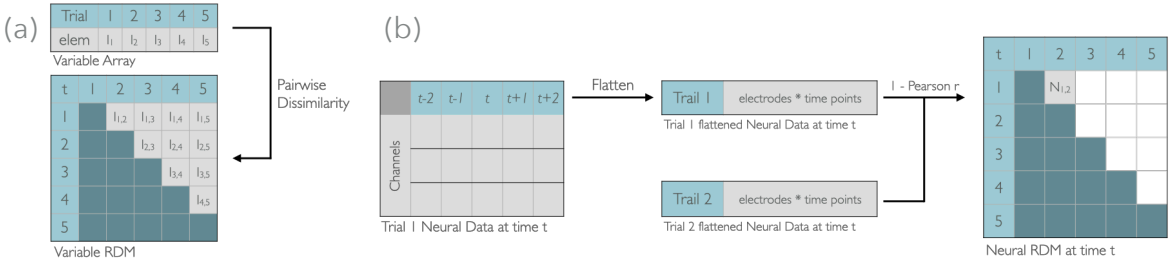

▸ Figure 1: Principles for obtaining the neural and variable RDMs. In the context of this paper, each trial has a unique stimulus. For demonstrative purposes, the figure only shows an RDM with five trials; the actual datasets analyzed in this thesis contain hundreds of trials, each with its unique stimulus. (a) The process for calculating a variable RDM. The variable array records the stimulus-specific property of interests (e.g., the word2vec embeddings of the words that approximate their semantic meanings). Pairwise dissimilarity value is calculated for all possible pairs of trials, and the values are assembled into a square RDM. Each cell records the dissimilarity value between the trials on the row and column of the RDM (e.g., cell $I_{1,2}$ represents dissimilarity between trial 1 and trial 2). The cells on the diagonal are always 0, and the RDM is symmetrical along this diagonal. To eliminate redundant information, only the upper triangular part of the RDM will be used for RSA. This upper part will be flattened into a long vector prior to analysis. (b) The process for calculating the neural RDM. The single-trial EEG recordings are arranged into the matrix on the left, with rows representing recording channels and the columns representing time points. Note that this matrix includes both the channel activation patterns at time $t$ and the channel activation patterns at the four neighboring time points. Hence the neural data of trial 1 at time $t$ is a matrix encapsulating the spatiotemporal activation patterns of the channels in a small time window. The matrix of this trial is then flattened into a vector, and the neural dissimilarity value between two trials at time $t$ is the Pearson correlation distance ($1 - Pearson\, r$) between their corresponding vectors (middle). The neural dissimilarity is then recorded in the cell of the neural RDM (right). Since RSA is repeated along the temporal dimension of the EEG recordings, a separate RDM is calculated for each time point.

recording activation values all channels. We can then calculate the pairwise neural dissimilarity value between any two vectors using correlation distance and construct a neural RDM relative to time point $t$. By repeating this process at all time points, we can obtain a *time series* of RDMs that encapsulates the temporal evolution of the neural representational geometry.

This naive approach outlined above demonstrates the basic principles for calculating neural RDMs. This approach only utilizes the spatial pattern confined to a particular time point, and the resulting RDM at each time point only captures the neural representational content in a tiny time window (usually 4 - 10ms, depending on the sampling rate). However, neural representations might also be encoded by the larger temporal pattern of brain activity. Therefore, a sliding time window approach can be employed to encapsulate the *spatiotemporal* activation pattern. This sliding time window approach combines both spatial and temporal patterns of neural activity by including the spatial activation patterns of several time points when assembling the vector for constructing neural RDM at a particular time point (Fig. 1b).

### Comparing RDMs

Once we have obtained psycholinguistics variable RDMs and a time series of neural RDMs for each subject, we can compare these variable RDMs to the neural RDMs. For each subject, an *RSA stream* repeatedly calculates Pearson partial correlation between the neural RDM at each time point and the variable RDM, producing a time series of coefficient R-values (Fig. 2). The result of this stream reveals the extent to which the variable representational geometry overlaps with the subject's neural representational geometries over time. This RSA stream is then repeated for each subject and each variable RDM of interest. To visualize the general predictive power of a specific variable on the neural data, we average the time series of its coefficient values across subjects.

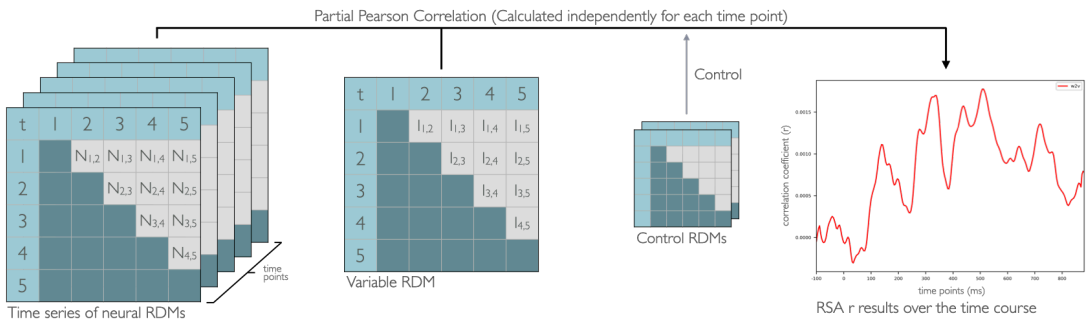

▸ Figure 2: Principles for comparing RDMs. For demonstrative purposes, only five trials are shown in this figure. The actual datasets analyzed in this thesis contain hundreds of trials, each with its unique stimulus. This figure demonstrates a single *RSA stream* that includes a single variable RDM and the time series of neural RDMs of a single subject. For the neural RDM at time point *t*, we calculate the Pearson partial correlation value between the upper triangular parts of the neural RDM and the variable RDM, controlling for the contributions of a group of control RDMs. The partial correlation will produce a correlation coefficient for time point *t*. This process is then repeated for all the neural RDMs across the time points, and the final product of this RSA stream is a time series of coefficient values, with time points recorded on the x-axis and the r values recorded on the y-axis.

However, the subject-specific results are retained for the statistical inference test.

### Statistical inference test

To test the statistical significance of the RSA results of a variable of interest, we use the cluster-based nonparametric statistical inference test designed by Maris and Oostenveld (2007). To summarize, the null hypothesis is a zero matrix whose dimensions match with those of the RSA results. This null hypothesis represents the situation where no relationship is detected between the variable RDM and the neural RDMs for all subjects. We first perform a paired t-test between the RSA results and the null hypothesis at each time point. Adjacent data points that exceed a pre-set uncorrected p-value threshold of 0.05 are bonded into a temporal cluster. The individual t-statistics within each cluster are summed to yield a cluster-level test statistics. We then randomly permute the RSA result and the null hypothesis on the subject level at each time point, and then calculate the cluster-level statistics. We take the largest cluster-level statistics (i.e., the summed t values) to build a null distribution after 1000 times of random permutation. Finally, we compare our observed cluster-level test statistics against this null distribution. All temporal clusters falling within the highest or lowest 2.5% of the distribution are considered significant.

# Aim 1: Testing the effectiveness of RSA on Single−trial EEG data

### Testing the effects of the number of trials and subjects on RSA results

The first aim of the thesis is to test the effectiveness of RSA on single-trial EEG data. We use the Kiloword dataset originally published by Dufau et al. (2015) for testing. The stimuli used in the dataset are 960 nouns, and the EEG data were collected from 55 participants while they performed a semantic categorization task (see Dufau et al., 2015 for details). All the analyses performed in this section follow the general processes for performing RSA that are introduced above. We build semantic variable RDMs from ratings of concreteness (Brysbaert et al., 2014), valence, arousal (Warriner et al., 2013), and lexical variable RDMs from orthographic Levenshtein distance (Yarkoni et al., 2008) and word length. While the Levenshtein distance RDM is obtained through its corresponding distance algorithm, all other variable RDMs are obtained

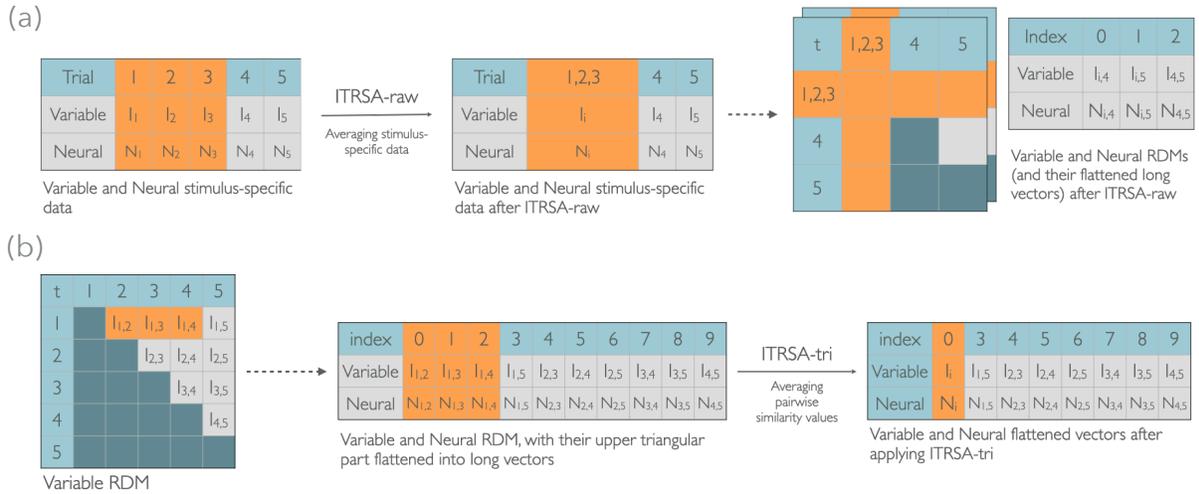

▸ Figure 3: Principles of the ITRSA-raw and ITRSA-tri. (a) ITRSA-raw interpolates stimulus-specific data. In this example, stimuli of trial 1, 2, & 3 have identical variable values, and the three trials are collapsed into one. For the neural data, neural recording epochs of trial $N_1$, $N_2$, & $N_3$ are averaged into a new neural recording epoch $N_i$. The effect of this interpolation on the RDMs and their flattened long vectors is shown on the right. (b) ITRSA-tri interpolates pairwise dissimilarity values. Since cells of RDMs are collapsed in this approach, the interpolation is actually carried out in the flattened long vectors of the RDMs. In this example, cells $I_{1,2}$, $I_{1,3}$, & $I_{1,3}$ have identical pairwise dissimilarity values in the variable RDM, so their corresponding columns in the flattened long vectors are collapsed into one. For the neural data, the neural dissimilarity values $N_{1,2}$, $N_{1,3}$, & $N_{1,3}$ are averaged into a new dissimilarity value $N_i$.

through the absolute difference function. Since some words in the stimuli set do not have arousal and valence ratings, only words trials are kept for this analysis. For all RSA streams, the lexical RDMs are used as control RDMs in the partial correlation analyses. To obtain the baseline performance, we first perform the analysis to all trials and subjects. To test the effect of the number of trials on the RSA results, we randomly select half of the trials (435) for all 55 subjects and apply RSA to the reduced stimulus set. To test the effect of subject size on the RSA results, we randomly select 20 subjects and apply RSA to the full stimulus set.

### RSA Interpolation

We also explore some interpolation techniques to actively raise the signal-to-noise ratio of the data to improve the sensitivity of single-trial EEG RSA. Here we want to introduce three general approaches: ITRSA-raw interpolates the stimulus-specific data, ITRSA-tri interpolates the pairwise dissimilarity values, and Grand-Average directly averages trials across subjects. The principles behind the first two approaches are illustrated in Fig 3. Here we apply the three approaches to the Kiloword dataset. In the ITRSA-raw approach, we interpolate trials that share the same concreteness/valence/arousal value prior to the construction of the variable and neural RDMs. In the ITRSA-tri approach, we directly interpolate the cells of the variable and neural RDMs if they share the same dissimilarity value relative to concreteness/valence/arousal. In the Grand-Average approach, we average recordings across subjects for each unique trial and construct a single time series of neural RDMs using the averaged neural data.

## Aim 2: Investigating the role of semantic meaning in emotional word processing

### Using semantic RDM to predict neural RDMs in the 500–800ms time window

The second aim of this thesis is to investigate the role of semantic meaning in emotional processing. For this purpose, we use the Emsingle dataset published in the Delaney-Busch et al. 2016 study. The stimuli used in the dataset are 468 English words. The words are equally split into three Part-of-speech (noun, adjective, adverb), and the three valence conditions (negative, neutral, positive) are fully crossed with the two arousal conditions (low, high). The lexical features are also matched between the valence and arousal conditions. In experiment 2, subjects view a unique word in each trial and are prompted to classify the word into three valence categories 800ms after stimulus onset, while their neural responses are recorded through EEG. All the analyses performed in this section follow the general processes for performing RSA that are introduced above. We use the word2vec word embeddings (Mikolov et al., 2013) of the words to obtain the semantic RDM, and the pairwise dissimilarity is defined as the cosine distance between word vectors. We use orthographic Levenshtein distance (Yarkoni et al., 2008) and word length to obtain the lexical RDMs, using the absolute difference function and the Levenshtein distance algorithm, respectively. In all RSA streams, the two lexical RDMs are taken as control. To test our first prediction, we include all the trials and apply the RSA streams on the word2vec RDM and the neural RDMs. We then perform the clustering test on the results in the 500-800ms time window to search for significant clusters. To test our second prediction, we first separate each subjects' trials into three valence conditions (negative, neutral, positive) according to their own judgments on the valence category of the words. We then build separate semantic and neural RDMs using trials from each condition and perform RSA streams independently. We perform the clustering permutation test on the results produced in each condition in the 500-800ms time window. Additionally, we also apply this cluster permutation test on the differences in results between every pairs of conditions to search for significant interactions.

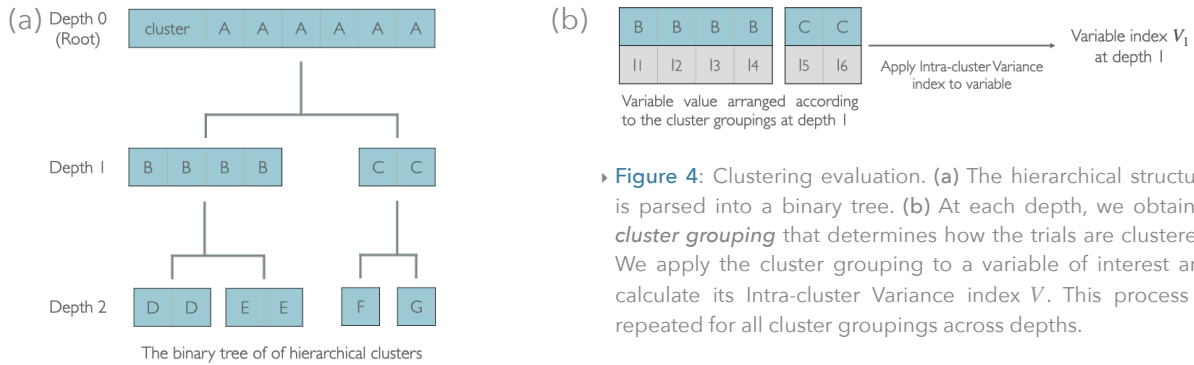

Figure 4: Clustering evaluation. (a) The hierarchical structure is parsed into a binary tree. (b) At each depth, we obtain a *cluster grouping* that determines how the trials are clustered. We apply the cluster grouping to a variable of interest and calculate its Intra-cluster Variance index $V$. This process is repeated for all cluster groupings across depths.

## Applying MDS and hierarchical clustering to the word2vec RDM

Additionally, we are interested in investigating whether common psycholinguistics variables like valence and arousal are captured by the word2vec RDM. In the literature, multidimensional scaling (MDS, Kruskal, 1978) and Hierarchical Clustering (Johnson, 1967) are usually applied to RDM to probe its representational content (e.g., Kriegeskorte et al. 2008, Nili et al. 2014, Kaneshiro et al., 2015). We apply MDS to the word2vec RDM to project the words onto a two dimensional space. We then correlate the words' projections on each of the two dimensions with the aforementioned psycholinguistics variables such as arousal, valence, concreteness, orthographic neighborhood (Heuven et al., 1998), word length, as well as log frequency (SUBTLEX-US; Brysbaert & New 2009). We also apply the hierarchical clustering algorithm implemented by the python Scikit-learn library (Pedregosa et al., 2011) to the word2vec RDM, using the Ward linkage criterion (Ward, 1963). We develop a novel method to analyze the extent to which each psycholinguistic variables influence the formation of the hierarchical clusters. Specifically, we parse the hierarchical clusters into a binary tree (Fig. 4a) and obtain a *clustering grouping* at each depth in the tree. We then use the clustering groupings to cluster the psycholinguistics variables and calculate their Intra-cluster Variance index values across the depths (Fig. 4b). This index measures the ratio of the intra-cluster variance and the variance of the dataset. Specifically, for a set of data $E$ of size $n_E$ that has been clustered into $k$ clusters, the Modified Intra-cluster Variance index $V$ is calculated as:

$$v = \sum_{q=1}^{k} \frac{n_q}{n_E} \times \frac{s_q^2}{s_E^2} \qquad s_q^2 = \frac{\sum (x_i - \bar{x})^2}{n_q - 1}$$

where $s_q^2$ is the variance of cluster $q$, $s_E^2$ is the variance of dataset $E$, and $n_q$ is the number of points in cluster $q$. A low index value indicates robust clustering property, and a variable with consistently low index values across all depth is considered to be an important principle in organizing the clusters.

# Results

## Aim 1: Testing the effectiveness of RSA on Single-trial EEG data

### The effects of the number of trials and subjects on RSA results

The effects of the number of trials and subjects on RSA are presented in Fig. 5a. While reducing the number of trials or subject increases the group-averaged r values, there are more fluctuations in the time series of the RSA values. Further, this increase in the magnitude of r values is at the expense of a reduced statistical power, since only a fraction of the significant temporal clusters found in the baseline condition is detected.

### The effects of the Interpolation on RSA results

The effects of the two interpolation approaches on RSA are presented in Fig. 5b. While both approaches successfully increase the r values of the result, their ability to raise the signal-to-noise ratio is ambiguous because the temporal patterns of the RSA results are altered by interpolation. Specifically, ITRSA-raw massively boosts the statistical power because the threshold of $p < 0.05$ leaves the results of all variables significant after stimulus onset. While the peaks of each result generally resemble the temporal pattern observed in the baseline result, the between-variable difference is smeared. ITRSA-tri also boosts the results' r values and keeps the temporal pattern relatively intact, but it seems to reduce the statistical power. Further, since the number of pairs of dissimilarity value being interpolated is vastly different between variables, this approach renders the between-variable comparison difficult. Finally, we also test the Grand-Average approach that averages trials across subjects. As we expected, this approach completely destroys the temporal patterns of the RSA time series we observed in the baseline results, and no significant cluster is detected. For this reason, its result is not shown in the figure.

## Aim 2: Investigating the role of semantic meaning in emotional word processing

### RSA results the 500-800ms time window

In this section, we will present the results obtained from the EmSingle dataset. First, we perform RSA on the word2vec RDM and the neural RDMs across subjects. Since the LPP effect reported by Delaney-Busch et al. (2016) only occurs in the 500-800ms time window, we restrict our statistical inference test to this time window. The result is shown in Fig. 6a. Our statistical test detects a single significant temporal cluster spanning

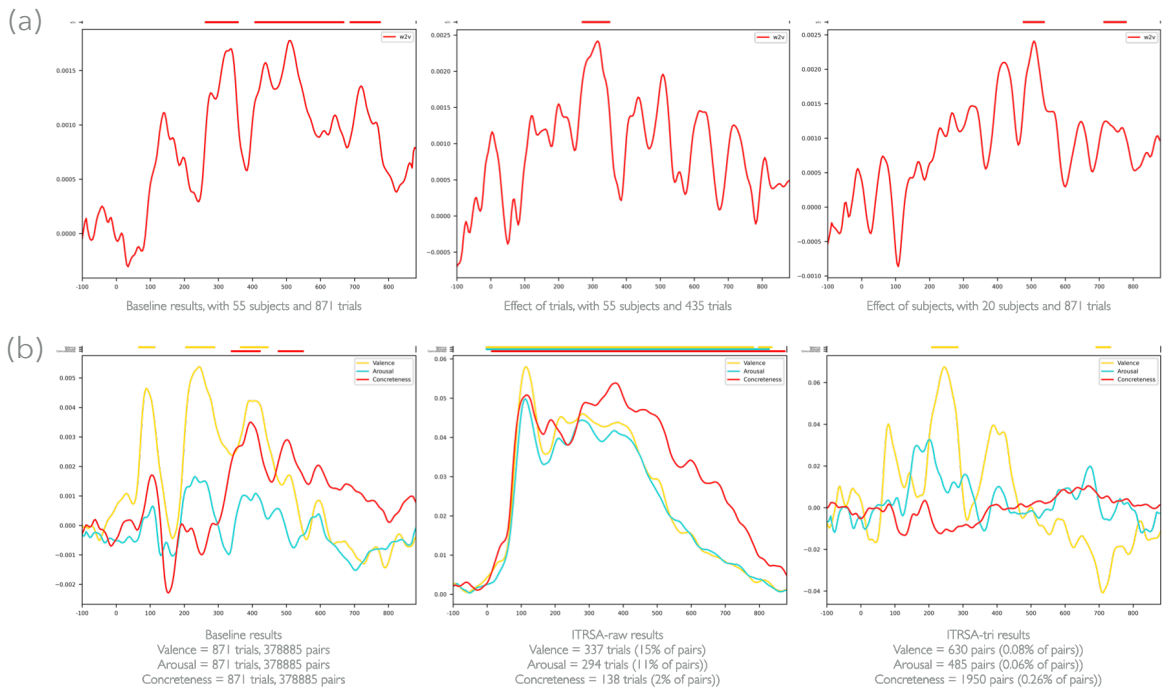

▸ Figure 5: RSA exploration results. In each plot, x-axis is the time (ms) and y-axis is the correlation coefficient value (r). Line segments above each plot indicate significant temporal clusters ($p < 0.05$) (a) The effect of the number of trials and subjects on RSA. Left: baseline result. Middle: result with reduced number of trials. Right: result with reduced number of subjects. (b) The effect of interpolation on RSA. Left: baseline results. Middle: results after ITRSA-raw. Left: results after ITRSA-tri.

the entire time window of interest. This result shows that the neurocognitive process operating in this time window is primarily engaging in semantic analysis. To test whether valence conditions modulate semantic processing in this time window, we apply RSA to each of the three valence conditions separately. The results are demonstrated in Fig 6b. Despite the noisy results, our statistical test detects significant temporal clusters in word2vec results for both the negative and the positive conditions in the 500-800ms time window. We also detect a significant interaction in the word2vec results between the negative and the neutral conditions in the 500-800ms time window. Together, these results indicate that valenced words trigger a deeper semantic analysis while the neutral words do not.

Common psycholinguistics variables encoded in the word2vec RDM

To investigate the semantic content captured in the word2vec RDM, we project the RDM onto a two-dimensional space using MDS and correlate each of the psycholinguistics variables with the words' projection values. The results presented in Fig. 6c and 6d indicate that valence (and to a lesser degree, concreteness) is an important dimension captured by word2vec RDM. We also analyze the hierarchical clustering results (Fig. 6d) obtained from the word2vec RDM using our novel approach (Fig. 4). The results presented in Fig. 6e indicate that valence is the most important organizing principle of the semantic clusters, followed by concreteness and log frequency. Both analyses produce highly consistent results, showing that the word2vec RDM captures most of the valence information and encapsulates some of the concreteness and log frequency information.

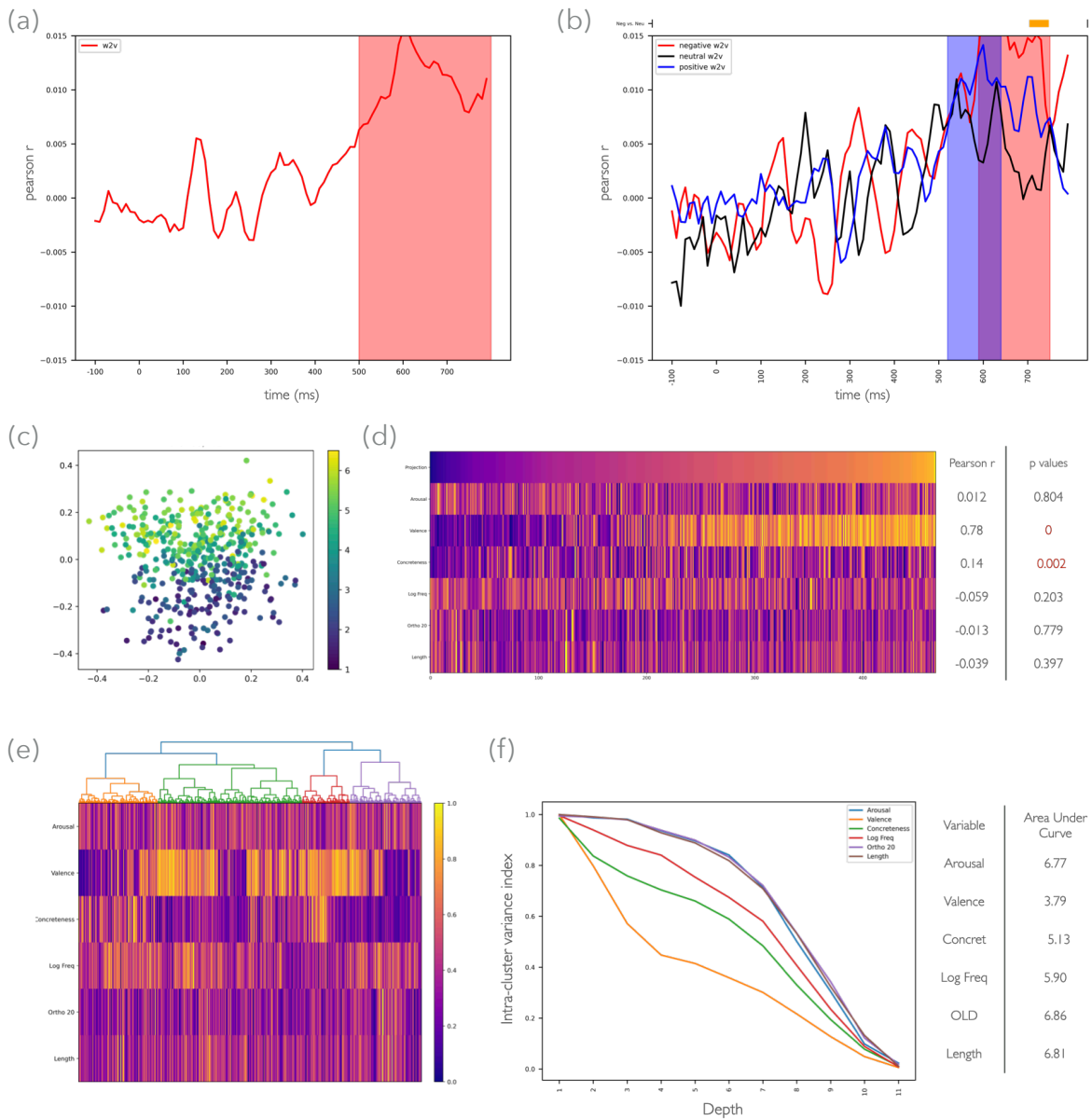

▸ Figure 6: EmSingle results. (a) The word2vec RSA results using all trials. The highlight region indicates a significant temporal cluster. (b) The word2vec RSA results on the negative, neutral, and positive valence condition respectively. Highlight regions indicate significant temporal clusters; line segments above the plot indicate significant interactions in the word2vec results between the negative and the neutral conditions based on individual subject's judgement. Interactions in the word2vec results between the negative/positive and positive/neutral conditions fail to reach significance. (c) The 2-dimensional MDS result of word2vec RDM, with words color-coded by their valence values. (d) Projections of the words onto the second dimension of MDS (y-axis). The first row indicates a perfect projection, and lower rows indicate the variables' values of each word reordered according to the words' projection. The rows indicate arousal, valence, concreteness, log frequency, orthographic Levenshtein distance, and word length, respectively. The list on the right indicates the pearson correlation coefficient between each reordered variable sequence and the perfect projection. A high coefficient between a variable sequence and the projection indicates that the variable's information is (partially) encoded in the word2vec RDM. (e) The Dendrogram of the hierarchical clustering applied to the word2vec RDM. Each row indicates the sequence of words' variable reordered according to the cluster sequence. (f) Each variable's Intra-cluster variance indexes across the entire hierarchy. A low index value indicates a robust clustering property. The list on the right records the area below each variable's curve. If a variable has a small area under the curve, then it is considered to be an important organizing principle for the clusters.

# Discussion

This thesis has two separated but related aims. Our first aim is to investigate the effectiveness of RSA on single-trial EEG data. To this end, we apply RSA to the Kiloword dataset. We test the effect of the number of trials and subjects, and experiment with three interpolation approaches. Our second aim is to use RSA to investigate the role of semantic meaning in emotional word processing. We apply RSA to the EmSingle dataset and test the two predictions made by our hypothesis. In this section, we will separate the two aims and discuss the results individually.

## Aim 1: Testing the effectiveness of RSA on Single-trial EEG data

### Effectiveness of RSA on EEG single-trial study

Noise is an ever-present challenge in univariate and multivariate analyses of neural data. Traditional RSA studies employ the condition-rich design (Kriegeskorte et al., 2008). In this design, neural recordings of several identical stimuli are collapsed into a single recording trial (condition) to increase the signal-to-noise ratio of the neural RDMs. However, due to the brain's extreme sensitivity to repeating language stimuli, typical psycholinguistics studies are limited to use unique stimuli for each trial. Applying RSA to datasets with unique stimuli means that the neural RDMs must be constructed from the noisy single-trial recording data. Such extreme noise levels raise a serious worry about RSA's utility on single-trial EEG data; this problem partially explains the lack of RSA studies in the psycholinguistics EEG literature. However, here we report that applying RSA to single-trial low-density EEG recordings can still produce meaningful and interpretable results. Specially, the baseline results in the Kiloword datasets yield significant temporal clusters in their expected time windows. Our results also indicate that a high number of trials and subjects are required to yield reliable RSA results.

In essence, the increase the number of trials and subjects leads to the increase the *reliability* of our measures. A higher number of trials can yield a more reliable neural representational geometry because the recorded neural response can better estimate the true neural response. Imagine that we are plotting the neural responses of each trial in a high-dimensional multivariate response space. The position of a trial on this response space is defined by its distance to all the other trials (and for this reason, neural RDMs reflect the neural representational geometries). However, the noise in recording data will shift the position of some trials in this response space. If such shift is extreme, the overall neural representable geometry will be seriously distorted. By increasing the number of neural recoding trials, we can make the overall recorded neural representational geometry more resistant to such anomalies, thereby rendering it a better estimate of the true neural representational geometry.

A higher number of subjects can also yield a more reliable neural representational geometry, although the exact reason behind its effectiveness is slightly different. Since the neural RDM only retains the second-order relationships between recording trials, it is invariant to rigid rotations and translations of the neural responses in the multivariate response space (Kriegeskorte & Diedrichsen, 2019). For this reason, neural RDMs can partially ignore the idiosyncrasies of the subjects' neural response and obtain an invariant neural representational geometry across subjects. In the ideal situation, subjects' neural RDMs will be identical if their neural activation patterns only reflect activities of the same set of neurocognitive processes triggered by the stimuli. But in reality, neural recordings of each subject also pick up activities of other neurocognitive processes that are irrelevant to the processing of the stimuli. Combined with recording noises, these idiosyncratic neural recordings can lead to a high level of dissimilarity between subjects' neural RDMs. However, as long as the task-irrelevant neurocognitive processes and recording noises are orthogonal to the experimental task, we can mitigate this problem by increasing the number of subjects. The more subjects we include in the analysis, the better we can delineate the set of invariant neurocognitive processes across subjects. Hence a high number of subjects is necessary to produce neural RDMs.

### Actively raising the signal-to-noise ratio through interpolation

In this thesis, we experimented with three interpolation approaches to actively raise the signal-to-noise ratio in the analysis. The evaluation of the true effectiveness of the approaches is beyond the scope of this study because we do not know the ground truth of the RSA values without running simulations. Here we only aim to discuss the essential concepts of the interpolation approaches. We wish that the ideas behind them can inspire future studies in this direction.

On a conceptual level, ITRSA-raw mimics the conditional-rich study design. Both the condition-rich design and ITRSA-raw assume that similar stimuli elicit similar neural responses and that the neural activity elicited by individual stimuli can be averaged to increase the signal-to-noise ratio of the neural data. However, in the condition-rich design, identical stimuli have the same measures in all aspects (e.g., valence, arousal). In contrast, in the ITRSA-raw interpolation, interpolated stimuli might only share identical values in one aspect as measured by a specific variable (e.g., stimuli might share the same valence rating but have different arousal ratings). ITRSA-raw essentially assumes a first-order isomorphic relationship between a particular variable of interest and neural responses. For example, by using ITRSA-raw to interpolate trials according to their valence rating, we assume that the difference between two stimuli in terms of their valence rating will reliably lead to a similar difference in their neural responses. This assumption does not always hold. In the above example, the neural responses of stimuli might be primarily modulated by their arousal but not valence ratings. Interpolating trials according to their valence values can potentially destroy meaningful neural responses. For this reason, ITRSA-raw should be employed only when there is a strong association between the variable of interests and the subject's neural responses.

ITRSA-tri is a novel attempt that operates on a more abstract level. On a conceptual level, RSA investigates the second-order relationships of a group of stimuli by measuring their pairwise dissimilarity; it reveals overlapping representational content if similar pairwise variable dissimilarity values in the variable RDM are paired with similar pairwise neural dissimilarity values. ITRSA-tri inherits this assumption and interpolates pairs of neural dissimilarity values if the pairs have identical variable dissimilarity values. Viewed this way, ITRSA-tri operates with a much weaker assumption than does ITRSA-raw. This approach is also more in line with the principles behind RSA since ITRSA-tri only assumes second-order isomorphic relationships between the variable and the neural data. ITRSA-tri can raise the signal-to-noise ratio because the random noises associated with the pairwise neural dissimilarity values are mitigated by averaging these neural dissimilarity values. Despite its promising potential on the conceptual level, however, ITRSA-tri only shows meager performance in our study. More detailed simulations are required to investigate the effectiveness of this approach.

Finally, the Grand-Average approach showed no improvement to the RSA. The stimulus-specific neural recording of each subject produces highly idiosyncratic signals. The grand-average approach assumes that a set of neurocognitive processes produces invariant recording *signals* (in contrast to *activation patterns*) in all subjects. Under this assumption, averaging the same trials across subjects can increase the signal-to-noise ratio. But this assumption does not always hold. While the grand-averaging approach in univariate analysis can sometimes produce invariant univariate components, the same approach might fail in multivariate analysis because it can destroy the neural spatial activation patterns that are crucial for differentiating different neural responses (Stokes et al., 2015). This is probably due to individual variability in their head shape, conductivity, or the location of recording electrodes. A more promising approach is to average neural data across subjects is to average the subjects' neural RDMs. As discussed in the previous section, since the neural RDM abstracts away information about activation levels of each recording channel and only retains the second-order relationship between different neural recordings, it automatically takes care of the idiosyncrasy of each subject's neural recordings. Hence the neural RDMs should be relatively invariant across subjects, and averaging these RDMs can increase the signal-to-noise ratio. In fact, this approach has been proposed as a measure of estimating the upper limit of the RSA values (Nili et al., 2014; Kriegeskorte et al. 2008). Our interests in the subject-rated valence category prevent us from using this approach, but we encourage future EEG studies to adopt this interpolation technique if possible.

# Aim 2: Investigating the role of semantic meaning in emotional word processing

## Emotional word processing in the 500–800ms time window is associated with additional semantic processing

ERP Literature on the LPP component indicates that neurocognitive processes in the 500-800ms time window after stimulus onset is related to emotional processing. Although previous studies reported that the motivational significance of the stimulus triggers the neurocognitive processes in this time window, little is known about the exact content being processed. We hypothesize that these neurocognitive processes are associated with semantic analysis on the motivationally significant stimuli. This hypothesis leads to two predictions about the RSA results: 1) The semantic RDM should be significantly correlated with the neural RDM in the 500-800ms time window; 2) After confining RSA to the trials in each of the three valence conditions (negative, neutral, positive), the semantic RDM should be significantly correlated with the neural RDM in the 500-800ms time window in the positive and the negative condition, but not in the neutral condition.

Although our hypothesis seems intuitive, it is not trivial. Specifically, it is not necessarily true that the emotional processing in the 500-800ms time window triggered by the motivationally significant stimulus is engaging in semantic processing. In fact, it is entirely possible that the brain switches off all semantic processing during this time window and is instead preparing autonomic motor responses. If this alternative hypothesis is true, then our results should disagree with the first prediction because the preparation of autonomic motor response should not produce neural activities that encode semantic information. Further, even though the brain might engage in additional semantic processing in this time window, it is entirely possible that the semantic processing is operating in parallel with the emotional processing and can be triggered by all kinds of stimuli. If this alternative hypothesis is true, then our result should agree with the first prediction but not the second prediction two. In particular, this alternative hypotheses requires that the neutral words should also trigger semantic processing. Therefore, these two predictions must be satisfied simultaneously to confirm our hypothesis.

Our results are consistent with the two predictions. When all trials are included in the analysis, we detect a robust cluster in the word2vec result that covers the entire time window of interest (500-800ms). This result indicates that the neurocognitive processes in this time window, in general, are associated with semantic processing. Further, we reapply the RSA analysis after separating the trials into three valence conditions and only detect significant clusters from the word2vec results in the positive and the negative condition in the time window. We also find a significant interaction in the word2vec results between the negative and the neutral condition. These results indicate that the neurocognitive processes in the 500-800ms time window are related to semantic processing only when the stimulus is motivationally significant. Together, our results support our hypothesis that the neurocognitive processes in the 500-800ms time window are associated with additional semantic analysis on the motivationally significant stimuli.

One explanation for this additional semantic analysis is that motivationally significant stimuli carry crucial information that demands careful judgments. While a motivationally significant visual imagery (i.e., the imagery of a tiger charging right at you) can be processed quickly, the same cannot be said about words that are presented on the screen in isolation. Words are naturally ambiguous in their meaning. Further, while motivationally significant words are often uttered differently (e.g., higher pitch, loud voice) in a natural setting, these extra-linguistics cues that can aid word processing are absent from the experiment. Therefore, the subjects likely need to engage in additional semantic analyses to determine the content of these emotional words.

Notably, Delaney-Busch et al. originally separated the three valence conditions according to a group of pre-rated valence values. However, they reported that the subjects' judgments sometimes disagree with the valence categories obtained from the pre-rated valence values. In this study, we choose to use each subjects' valence judgment to separate their trials into the three conditions. This is because the valence content (and in this context, the motivational significance) of a word is dependent on the subject's past experiences. For example, the word "firework" can carry a positive emotional content but might be interpreted to be negative if fireworks burned the subject in the past. In this way, we can make sure that the positive and negative conditions really contain motivationally significant stimuli.

Interestingly, our inability to detect a significant interaction in the word2vec results between the positive and the neutral condition seems to suggest the existence of a negativity bias. In general, a negativity bias is thought to reflect our inherent attentional bias toward stimuli with negative emotional content due to their association with the potential threat (Carretié, 2014). It is possible that we only observe a significant interaction in the word2vec results between the positive and the neutral condition because we devote extra resources to analyze the semantic content of negatively valenced words. However, given the fact that each valence condition contains less than 200 words, this asymmetry in the interaction effect might be attributed to the low statistical power. Therefore, the results presented here are ambiguous and cannot be used as evidence supporting the existence of a negativity bias.

### Valence is the most salient semantic dimension encoded in word2vec RDM

Since semantic RDM constructed from DSM can significantly predict neural activation patterns, an additional question we want to explore here is the content of the semantic meaning captured by the DSM. Previous literature on DSMs reveals that they can encode rich semantic information. Common psycholinguistics variable information such as arousal and valence may be already encoded in the DSM-based semantic RDMs. For this reason, we apply MDS and Hierarchical Clustering to the word2vec RDM and investigate whether common psycholinguistics variable values can be reconstructed from the semantic representational geometry. Much to our surprise, we find that the arousal information is not present in the word2vec RDM even though it is considered one of the two main dimensions of the affective content of words (Delaney-Busch et al., 2016). Further, although we expect the valence information to be partially encoded in the word2vec RDM, its saliency greatly exceeds our expectations. In particular, we are surprised to observe that the words projection onto the second dimension of the MDS result is almost perfectly correlated with their valence values. Our hierarchical clustering analysis also reveals that valence is a dominating organizing principle for the formation of semantic clusters. These results suggests that valence values can be decodable from the word2vec RDM and that valence is perhaps one of the most salient emergent dimensions from the semantic representation captured by word2vec.

The fact that DSM can easily capture valence information suggests that valence can be easily and consistently represented using distribution statistics of words. This insight raises interesting questions regarding the nature of valence. For example, it is possible that valence is not a particular semantic feature but instead an abstract semantic dimension emergent from the interaction of a large group of semantic features. This view might serve as an additional explanation of our observation that emotional word processing is associated with additional semantic processing. However, the nature of valence information is far beyond the scope of this paper, and the additional analyses we presented here are highly explorative in their nature. We hope the results presented here can inspire future research in this direction.

# Data Availability

Here we report a new python-based RSA toolbox specialized for single-trial EEG dataset (https://github.com/tedchengf/EEG_RSA). All the analyses in this paper are performed using this toolbox.